\documentclass[12pt]{article}
\usepackage{amsmath,mathrsfs,amsthm}
\usepackage{amssymb}
\usepackage{graphicx,psfrag,epsf}
\usepackage{enumerate}
\usepackage{bm}
\usepackage{multirow}
\usepackage{color}
\usepackage{natbib}
\usepackage{url} % not crucial - just used below for the URL 
\usepackage{color}
\usepackage{colortbl}
\usepackage{mathtools}
\usepackage{enumitem}
\usepackage{subfigure}
\usepackage{ctable}
\usepackage{array,longtable}
\usepackage[normalem]{ulem} % \sout

\usepackage{dcolumn}
\newcolumntype{d}[1]{D{.}{.}{#1}}

\makeatletter
\def\old@comma{,}
\catcode`\,=13
\def,{%
  \ifmmode%
    \old@comma\discretionary{}{}{}%
  \else%
    \old@comma%
  \fi%
}
\makeatother

\newcommand{\blind}{1}
\def\bx{\boldsymbol{x}}
\def\bX{\boldsymbol{X}}

\newcommand{\vecU}{\boldsymbol{U}}

\def\bmu{\boldsymbol{\mu}}
\def\bgamma{\boldsymbol{\gamma}}

\def\btheta{\boldsymbol{\theta}}
\def\bSigma{\boldsymbol{\Sigma}}

\newcommand{\bzero}{\boldsymbol{0}}

\def\bI{\boldsymbol{I}}

\def\bgamma{\boldsymbol{\gamma}}

\newcommand{\real}{I\hspace{-1.0mm}R}

\begin{document}

\def\spacingset#1{\renewcommand{\baselinestretch}%
{#1}\small\normalsize} \spacingset{1.3}

\if1\blind
{
  \title{\bf The generalized hyperbolic family and automatic model selection through the multiple-choice LASSO}
  \author{Luca Bagnato    \hspace{.2cm}\\
    Department of Economic and Social Sciences\\
    Catholic University of the Sacred Heart - Piacenza\\
    Alessio Farcomeni
    %\thanks{
    %The authors gratefully acknowledge \textit{please remember to
    %list all relevant funding sources in the unblinded version}}
    \hspace{.2cm}\\
    Department of Economics and Finance\\University of Rome ``Tor Vergata''\\
    and \\
    Antonio Punzo \\
    Department of Economics and Business\\ University of Catania }
  \date{}
  \maketitle
} \fi

\if0\blind
{
\title{\bf The generalized hyperbolic family and automatic model selection through the multiple-choice LASSO}
  \author{}
  \date{}
  \maketitle
} \fi

\bigskip
\begin{abstract}
  We revisit the generalized hyperbolic (GH) distribution and its nested models. 
	These include widely used parametric choices like the multivariate normal, skew-$t$, Laplace, and several others. 
	We also introduce the multiple-choice LASSO, a novel penalised method for choosing among alternative constraints on the same parameter.
  A hierarchical multiple-choice LASSO penalised likelihood is optimised to perform simultaneous model selection and inference within the GH family.
  We illustrate our approach through a simulation study.
  The methodology proposed in this paper has been implemented in \textsf{R} functions which are available as supplementary material. 
\end{abstract}

\noindent%
\vfill

\spacingset{1.5} % DON'T change the spacing!

\section{Introduction}
\label{intro}

As stated by \citet{cox:90}, ``choice of an appropriate family of distributions may be the most challenging phase of analysis''. Researchers always face a trade-off between goodness of fit and simplicity of the distributional assumptions.
A particularly convenient family is provided by the generalized hyperbolic (GH) distribution \citep[e.g.,][]{mcne:05}. It has flexible tails, spanning from Gaussian to exponential tails.
Applications of the GH family are widespread (e.g., \citealp{eber:kell:95,mcne:05}), and more importantly, the family contains as special cases several widely used parametric distributions. 
A contribution of this work indeed is that we outline a precise taxonomy of the GH family and its many nested models. 
The main novelty with respect to previous works is that we do not compare the GH and alternatives by separately fitting each model,
but we specify a unified penalised likelihood framework that successfully performs simultaneous parameter estimation and model choice. 

To proceed in this direction, we introduce the multiple-choice LASSO, a new type of LASSO penalty.
Indeed, LASSO-type penalties \citep{tibs:96} are commonly used to shrink parameters to a single specific value (typically, zero).
Nested models within the GH family are selected by fixing certain shape parameters at one of the different alternative values. 
The multiple-choice LASSO is devised precisely for this purpose: to allow shrinkage of the same parameter towards one of several alternative values.
To restrict the possible choices, we will also build on the hierarchical LASSO (as introduced by \citealp{bien:tayl:tibs:13}, see also \citealp{lim:hast:15}) so that certain constraints can be activated only conditionally. 

The rest of the paper is as follows: in the next section, we review the GH distribution and provide a map of its nested models.
After reviewing LASSO and hierarchical LASSO we then introduce the multiple-choice LASSO.
In Section~\ref{sec:penalties} we use the hierarchical and multiple-choice LASSO to define penalised objective functions that can yield any model within the GH family, and describe how to optimise those in Section~\ref{sec:Penalised maximum likelihood estimation}.
In Section~\ref{sec:Simulation study} we illustrate through a brief simulation study.
Some concluding remarks are given in Section~\ref{sec:conclusions}. 

The methodology proposed in this paper has been implemented in \textsf{R} \citep{R:2020} functions which are available as supplementary material. 

\section{Setup}

\subsection{The generalised hyperbolic distribution and its special cases}
\label{sec:GH distribution}

%From \cite{mcne:05}, 
The joint probability density function of a $d$-variate random variable $\bX$ following the generalised hyperbolic (GH) distribution
%, in symbols $\mathcal{GH}_d\left(\bmu,\bSigma,\bgamma,\lambda,\chi,\psi\right)$, 
%with parameters $\bmu \in \real^d$, $\bSigma\in \real^{d\times d}$, $\bgamma \in \real^d$, $\lambda \in \real^d$, $\chi\in \real^+$, and $\psi\in \real^+$ 
can be written as
\begin{align}
f\left(\bx;\btheta\right)=&\frac{\exp\left[(\bx-\bmu)'\bSigma^{-1}\bgamma \right]}{(2\pi)^{\frac{d}{2}}| \bSigma |^{\frac{1}{2}}K_{\lambda} \left(\sqrt{\chi\psi}\right)}  \left[\frac{\chi + \delta(\bx;\bmu,\bSigma)}{\psi + \rho(\bgamma,\bSigma) }\right]^{\frac{\lambda-\frac{d}{2}}{2}} 
%\nonumber \\
%& \times  
K_{\lambda-\frac{d}{2}}\left(\sqrt{\left[\chi + \delta(\bx;\bmu,\bSigma) \right] \left[\psi + \rho(\bgamma,\bSigma) \right]}\right),
\label{eq:gh}
\end{align}
where $\bmu\in\real^d$ is the location parameter, $\bSigma$ is a $d\times d$ scale matrix, such that $\left|\bSigma\right| = 1$ for identifiability purposes \citep[see][for details]{mcne:05}, $\bgamma\in\real^d$ is the skewness parameter, $\lambda\in\real$ is the index parameter, and $\chi,\psi>0$ are concentration parameters; compactly, we adopt the notation $\bX \sim \mathcal{GH}_d\left(\bmu,\bSigma,\bgamma,\lambda,\chi,\psi\right)$.
In \eqref{eq:gh}, $\btheta=\left\{\bmu,\bSigma,\bgamma,\lambda,\chi,\psi\right\}$ contains all the parameters of the model, $\delta(\bx;\bmu,\bSigma)=(\bx-\bmu)'\bSigma^{-1}\left(\bx-\bmu\right)$ is the squared Mahalanobis distance between $\bx$ and $\bmu$ (with covariance matrix $\bSigma$), $\rho(\bgamma,\bSigma)=\bgamma'\bSigma^{-1}\bgamma$, and $K_{\lambda}$ is the modified Bessel function of the third kind with index $\lambda$.

It is of practical importance to note that $\bX \sim \mathcal{GH}_d\left(\bmu,\bSigma,\bgamma,\lambda,\chi,\psi\right)$ has the normal mean-variance mixture (NMVM) representation
\begin{equation}
\bX=\bmu+W\bgamma+\sqrt{W}\vecU,
\label{eq:vmm}
\end{equation}
where $W$ has a generalised inverse Gaussian (GIG) distribution, in symbols $W \sim \mathcal{GIG}\left(\lambda,\chi,\psi\right)$ (see \appendixname~\ref{app:Generalised inverse Gaussian distribution}), and $\vecU\sim\mathcal{N}_d\left(\bzero,\bSigma\right)$, where $\mathcal{N}_d\left(\bmu,\bSigma\right)$ denotes a $d$-variate normal distribution with mean $\bmu$ and covariance matrix $\bSigma$.
As a related alternative, we can refer to the following hierarchical representation of $\bX \sim \mathcal{GH}_d\left(\bmu,\bSigma,\bgamma,\lambda,\chi,\psi\right)$ as 
%The representation in \eqref{eq:vmm} can be equivalently written as 
\begin{align}
	W &\sim \mathcal{GIG}\left(\lambda,\chi,\psi\right) \nonumber \\
	\bX|W=w &\sim \mathcal{N}_d\left(\bmu + w \bgamma, w\bSigma\right), \label{eq:hierarchical GH representation}
\end{align}
where $w$ is a realization of $W$.
The hierarchical representation in \eqref{eq:hierarchical GH representation} is useful for random data generation and for the implementation of the ECME algorithm discussed in~Section~\ref{sec:Penalised maximum likelihood estimation}.

\begin{figure}[!htb]
\centering
\includegraphics[width=\textwidth]{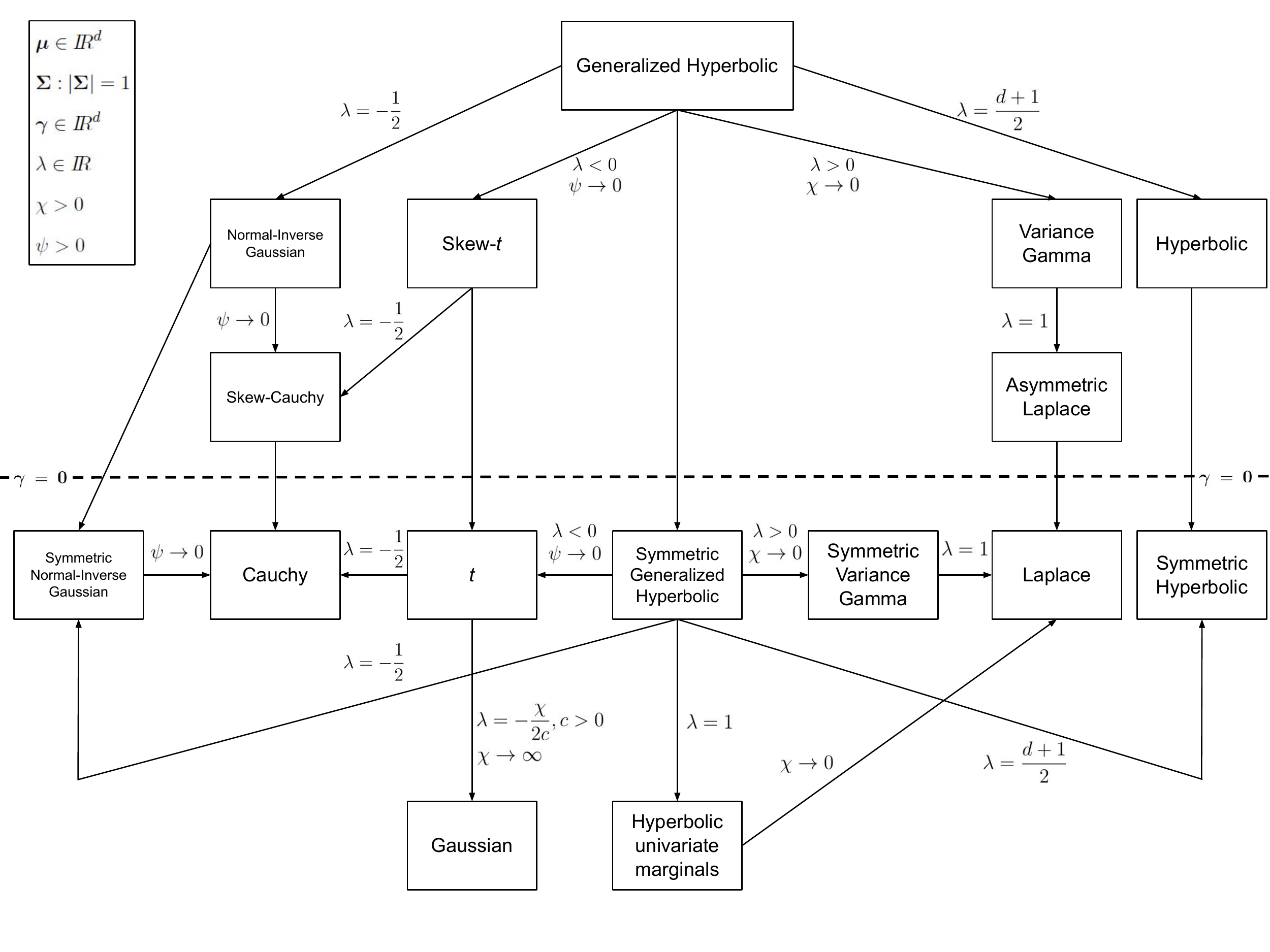}
\caption{Hierarchy of the special and limiting cases of the GH distribution in terms of $\bgamma$, $\lambda$, $\chi$ and $\psi$.
On the top-left corner, a recap on the values the GH-parameters can assume is provided.} 
\label{mcneil2}
\end{figure}

\figurename~\ref{mcneil2} gives a hierarchical representation of all the existing models the GH distribution nests as special or limiting cases by varying the values/ranges of $\bgamma$, $\lambda$, $\chi$, and $\psi$.
Such a hierarchy is easily derived by using the representation of the GH distribution given in \eqref{eq:vmm}.
\appendixname~\ref{app:Special and limiting cases of the GH distribution} illustrates how to obtain some of these special and limiting cases, those we believe are more difficult to be derived and about which there is more confusion in the literature due to the use of different identifiability constraints.
On the left/right of \figurename~\ref{mcneil2} we have the models related to negative/positive values of $\lambda$.
Instead, on the bottom (below the dashed line) we have the symmetric models (those with $\gamma=0$); as we can see, the symmetric counterpart of each model on the top is available. 
%each model on the top possesses its symmetric counterpart.
%\textcolor{blue}{It is important to note that all these models can be obtained thanks to the parametrization of the GH distribution we give in \eqref{eq:gh}.} 
The diagram in \figurename~\ref{mcneil2} can be considered as a contribution of this paper. It provides, for the first time to our knowledge, a complete and organised taxonomy of all the models nested within the GH family.

Summarising we have: 2 possibilities for $\bgamma$ ($\bgamma$ free or $\bgamma=\bzero$), 6 possibilities for $\lambda$ ($\lambda \rightarrow -\infty$, $\lambda<0$, $\lambda=-1/2$, $\lambda=\left(d+1\right)/2$, $\lambda=1$ or $\lambda>0$), 3 possibilities for $\chi$ ($\chi$ free, $\chi\rightarrow 0$ or $\chi\rightarrow \infty$), and 2 possibilities for $\psi$ ($\psi$ free and $\psi\rightarrow 0$).
Combining all these possibilities would generate $2\cdot 6\cdot 3\cdot 2=72$ models.
However, many of them are not of practical interest.
Just as two examples, the combination $\left\{\bgamma=\bzero,\lambda<0,\chi\rightarrow 0,\psi\rightarrow 0\right\}$ would generate a degenerate $t$ distribution on $\bmu$, while the combination $\left\{\bgamma=\bzero,\lambda=1,\chi\rightarrow 0,\psi\rightarrow 0\right\}$ would generate a degenerate Laplace distribution on $\bmu$.

\subsection{Preliminaries about LASSO and hierarchical LASSO}
\label{subsec: Preliminaries about LASSO and hierarchical LASSO}

Suppose to be interested to a particular configuration/value of $\btheta$, say $\btheta_0$.
The LASSO (Least Absolute Shrinkage and Selection Operator) involves specification of an $L_1$ penalty for (possibly, a subset of) the parameter vector $\btheta$, so that the estimate $\hat\btheta$ is exactly equal to $\btheta_0$ if the likelihood at $\btheta_0$ is not too far from the maximum.
More formally, given a random sample $S_n=\left\{\bx_i; i=1,\ldots,n\right\}$ (observed data) from $\bX \sim \mathcal{GH}_d\left(\bmu,\bSigma,\bgamma,\lambda,\chi,\psi\right)$, estimation proceeds through optimisation of the penalised log-likelihood
%More formally, let $\bx_i$, $i=1,\ldots,n$, denote i.i.d.~replicates of a $d$-variate random variable $\bX$. 
%Estimation proceeds through optimization of the penalised log-likelihood
\begin{equation}
  \label{plik}
  \sum_{i=1}^n \log\left[f\left(\bx_i;\btheta\right)\right] - P_h\left(\btheta\right) 
\end{equation}
for an appropriate penalty function $P_h\left(\btheta\right)$, with $f\left(\cdot;\btheta\right)$ being defined in \eqref{eq:gh}. 
In classical LASSO, $P_h\left(\btheta\right) = h ||\btheta-\btheta_0||_{L_1}$, where $||\cdot||_{L_1}$ indicates the $L_1$-norm (the sum of absolute values) and $h>0$ is a fixed penalty parameter.
In linear models, often times $\btheta_0=\bzero$. 

The resulting estimator is less efficient than the MLE, but superefficient at $\btheta_0$ (see, e.g., \citealp{wu:zhou:19} and references therein). 
It is well known that any superefficient estimator may improve efficient estimators at most on a subset of the parameter space of zero Lebesgue measure.

In our work we will also make use of the hierarchical LASSO \citep{bien:tayl:tibs:13}, which is devised for structured sparsity: some constraints can be activated
only if others are simultaneously active. 
%Let $\theta_c$ and $\theta_d$ be two elements of $\btheta$.
Without loss of generality assume we allow $\theta_c=0$ only if $\theta_d=0$, with $\theta_c$ and $\theta_d$ being two elements of $\btheta$. 
This can be obtained expressing
$$
P_h\left(\btheta\right) = h \left[|\theta_d| + \frac{\max(|\theta_c|,|\theta_d|)}{2}\right].
$$
In words, some shrinkage for $\theta_c$ is allowed if $|\theta_c|>|\theta_d|$,
but the constraint on $|\theta_c|$ can be exactly activated only as soon
as $\theta_d=0$; see \cite{bien:tayl:tibs:13} on this point.

\subsection{The multiple-choice LASSO}
\label{mcLASSO}

We introduce in this section the multiple-choice LASSO, which can be used to enforce one of several constraints on the same parameter.
For simplicity assume we have a one-dimensional parameter $\theta$ and several possible constraints on it, i.e., we require superefficiency not only
at a single point $\theta_0$ in the parameter space, but at a finite collection of points $\left\{\theta_1,\ldots,\theta_C\right\}$.
%For simplicity assume we have several possible constraints on the parameter $\theta\in\btheta$, i.e., we require superefficiency not only at a single point $\theta_0$ in the parameter space, but at a finite collection of points $\theta_0=\left(\theta_1,\ldots,\theta_C\right)'$.
Our proposal is to specify
\begin{equation}
  \label{eq:mcLASSO}
  P_h\left(\theta\right) = h \min\left(|\theta-\theta_1|,|\theta-\theta_2|,\ldots,|\theta-\theta_C|\right).
\end{equation}
In words, only the smallest among all possible $L_1$ norms contribute to the penalty. 
The idea is that if the MLE is close enough to $\theta_j$ for some $j=1,\ldots,C$, then $\hat\theta=\theta_j$ as the remaining $L_1$ norms are simply ignored due to the minimum operator.

%The multiple-choice LASSO can be also combined with the hierarchical LASSO, as we show in the next section. E.g., without loss of generality suppose one wants to activate
%any constraint on $\theta_c$ (e.g., that is it either $\theta_{1c}$, $\theta_{2c}$, up to $\theta_{Cc}$) only if either $\theta_d=\theta_{1d}$, $\theta_d=\theta_{2d}$, up 
%to $\theta_d=\theta_{Dd}$), then one can specify
%\begin{eqnarray*}
%  P(\theta) &=& w \min(|\theta_d-\theta_{1d}|,|\theta_d-\theta_{2d}|,\ldots,|\theta_d-\theta_{Dd}|)+\\
%  && + w \max( \min(|\theta_d-\theta_{1d}|,|\theta_d-\theta_{2d}|,\ldots,|\theta_d-\theta_{Dd}|),
%\min(|\theta_c-\theta_{1c}|,|\theta_c-\theta_{2c}|,\ldots,|\theta_c-\theta_{Cc}|)) / 2.
%\end{eqnarray*}

\begin{figure}[!ht]
\centering
\subfigure[LASSO \label{lasso}]
{\resizebox{0.48\textwidth}{!}{\includegraphics{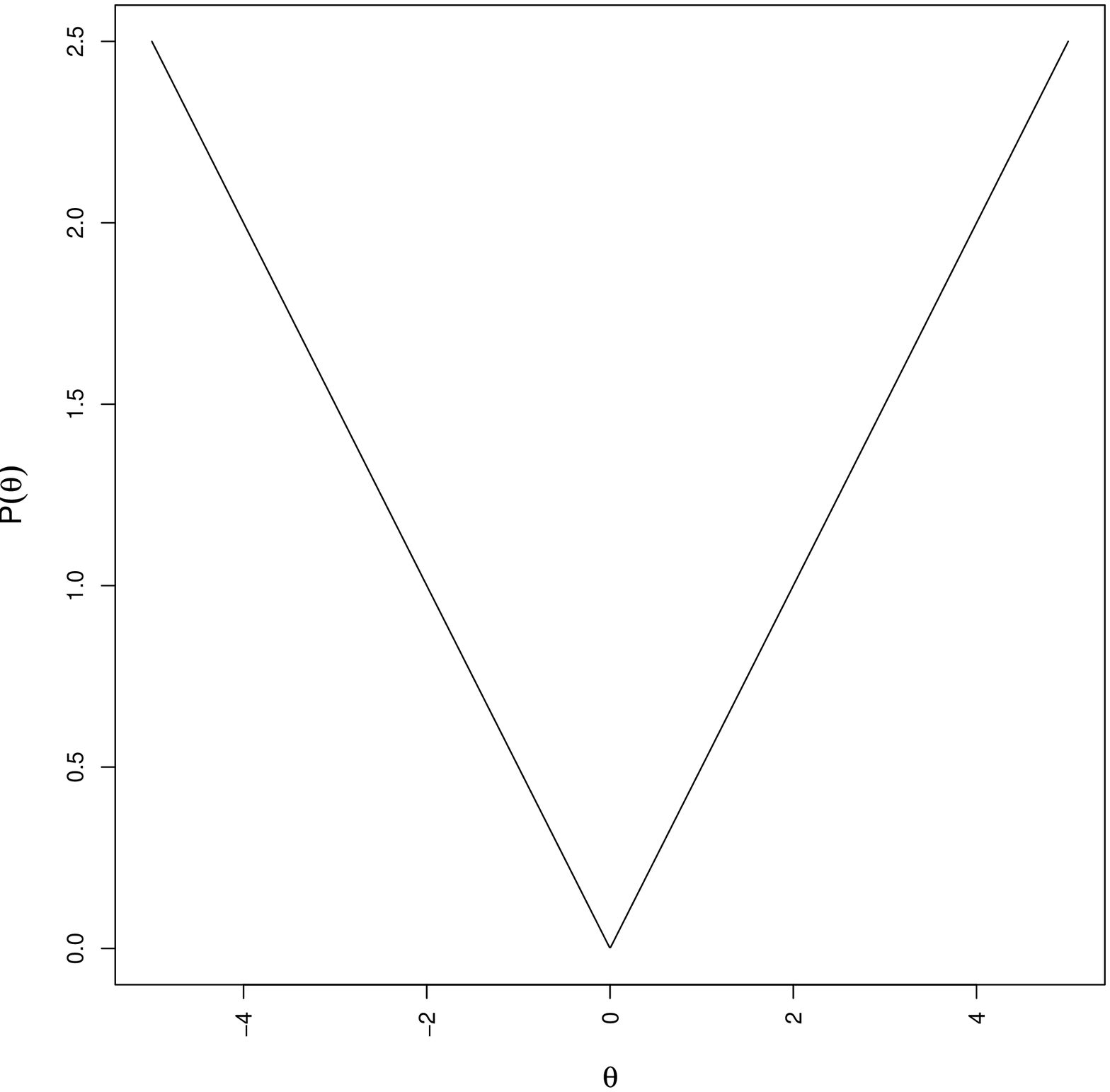}}}
\subfigure[Multiple-choice LASSO \label{multlasso}]
{\resizebox{0.48\textwidth}{!}{\includegraphics{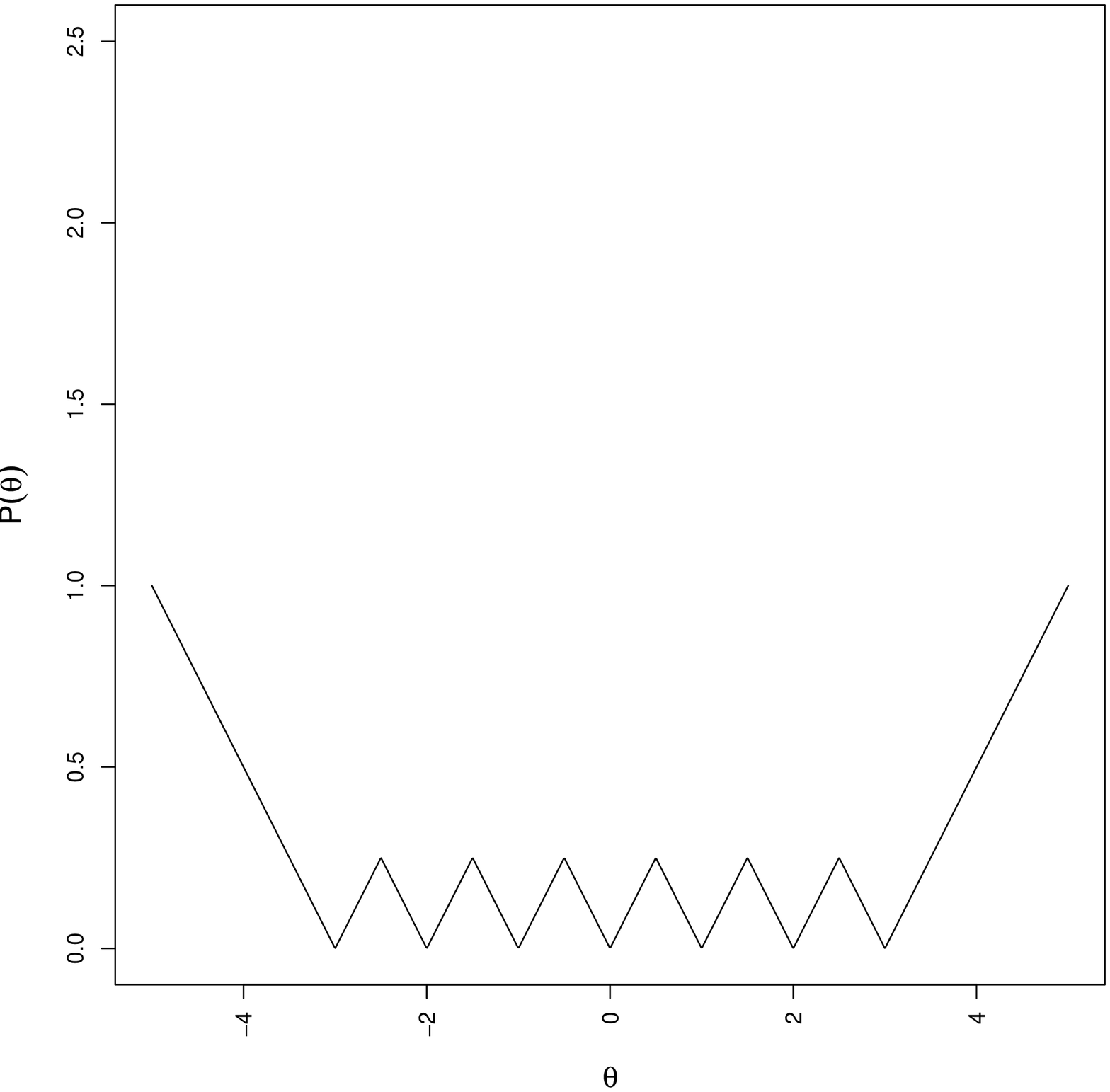}}}
\caption{
\footnotesize 
The penalty function for LASSO (left panel), with $\theta_0=0$; and the penalty function for multiple-choice LASSO (right panel), with $\theta_0\in\left\{-3,-2,-1,0,1,2,3\right\}$.
\label{mcLASSill}
}
\end{figure}

For illustration, in \figurename~\ref{lasso}--\ref{multlasso} we show the penalty function for LASSO and multiple-choice LASSO, respectively, for a one-dimensional problem with $h=0.5$ in both cases. 
For the LASSO we set $\theta_0=0$, while for multiple-choice LASSO we set $\theta_0\in\left\{-3,-2,-1,0,1,2,3\right\}$. 
The sawtooth shape of the penalty function for the multiple-choice LASSO is what allows objective functions to be optimised exactly at $\theta_j$, $j=1,\ldots,C$.
%\begin{figure}[!ht]
%\centering
%\includegraphics[width=0.8\textwidth]{estimationPicture.pdf}
%\caption{The penalty function for LASSO (left panel), with $\theta_0=0$; and the penalty function for multiple-choice LASSO (right panel), with $\theta_0\in\left\{-3,-2,-1,0,1,2,3\right\}$.} \label{mcLASSill0}
%\end{figure}

The resulting penalised objective function is clearly non-convex.
While in some cases specific algorithms might be exploited to optimise it, since the parameter space is low dimensional in our context, we propose to simply use
a numerical method like the Constrained Optimisation BY Linear Approximation (COBYLA) algorithm \citep{powe:94}. 

\section{Shape detection through penalised likelihood maximization}
\label{sec:penalties} 

As discussed at the end of Section~\ref{sec:GH distribution}, all possible combinations of the discussed constraints on the parameters $\bgamma$, $\lambda$, $\chi$, and $\psi$ would lead to 72 parametric distributions, nested within the GH distribution.
Of these, only 16 have a clear interpretation as outlined in Section~\ref{sec:GH distribution} and \figurename~\ref{mcneil2}.

In the following, we show how to specify a multiple-choice LASSO-type penalised likelihood function which can possibly lead to any of the 72 models nested in the GH distribution. 
We then specify a multiple-choice hierarchical LASSO-type penalised likelihood which restricts the possible solutions only to the sixteen models in \figurename~\ref{mcneil2}.

The penalised likelihood specification is as in \eqref{plik}. 
A simple way to proceed is to specify $P_h\left(\bgamma,\lambda,\chi,\psi\right)$ as a multiple-choice LASSO penalty of the kind 
\begin{equation}
  \label{60modelli}
  P_h\left(\bgamma,\lambda,\chi,\psi\right) = h \left\{ \min\left[\left|\lambda-\frac{d+1}{2}\right|, \left|\lambda+\frac{1}{2}\right|, |\lambda-1|, I(\lambda<0)\left|\frac{1}{\lambda}\right|\right] + \min\left(|\chi|,\left|\frac{1}{\chi}\right|\right) + |\psi| + \left\|\bgamma\right\|_{L_2} \right\}. 
\end{equation}
We use here a penalty on $\left\|\bgamma\right\|_{L_2}$ to constrain all $d$ elements of $\bgamma$ to be zero, in the spirit of group LASSO (see, e.g., \citealp{yuan:lin:06} and \citealp{lim:hast:15}). 
In case $\lambda\to -\infty$ and $\chi\to\infty$, define $c=-\chi/2\lambda$ as scale parameter of the resulting
Gaussian distribution. Note that the constraint $|1/\lambda|$ is satisfied by $\lambda\to\pm\infty$.

Penalty \eqref{60modelli} will allow the user to select any of the 72 possible parametric distributions obtained through appropriate constraints.
Many of these models might fit well, but do not have a direct interpretation. 
In order to restrict the list of possible models to the sixteen ones listed in \figurename~\ref{mcneil2} we must exclude several possible combinations
of constraints on the parameters. 
To this end, we combine the hierarchical LASSO and the multiple-choice LASSO frameworks and specify the penalty as 
\begin{align}
  \label{16modelli}
  P_h\left(\bgamma,\lambda,\chi,\psi\right) = & 
	h \left\{ 
	\frac{\left\|\bgamma\right\|_{L_2}}{\sqrt{d}}  
	+ I\left(\lambda\leq 0\right) 
	\min\left[\left|\lambda + \frac{1}{2}\right| + \frac{1}{2}\max\left(\left|\lambda+\frac{1}{2}\right|,\left|\psi\right|\right),\right.\right.\\
	&\nonumber\left.  |\psi| + \frac{1}{2}\max\left( \left|\lambda+\frac{1}{2}\right| , \left|\psi\right| \right), \frac{1}{4}\max\left( \frac{\left\|\bgamma\right\|_{L_2}}{\sqrt{d}},\left|\frac{1}{\lambda}\right|,|\psi|,\left|\frac{1}{\chi}\right|\right)\right] + \\ 
  & \nonumber \left. +I(\lambda>0) \min\left[\left|\lambda-\frac{d+1}{2}\right|, 
   |\chi|+\frac{1}{2}\max(|\lambda-1|,|\chi|), \frac{1}{2}\max\left(\frac{\left\|\bgamma\right\|_{L_2}}{\sqrt{d}},|\lambda-1|\right)\right]
	\right\},
\end{align} 
where $I\left(A\right)$ denotes the indicator function of $A \subseteq \real$ and $h>0$ is a penalty parameter. In the expression above we divide by $\sqrt{d}$ to normalize
the $L_2$ norm with respect to the number of elements of the vector involved. 

To fix the ideas we discuss how the GH and Gaussian models are obtained. 
If the MLE is far from any of the special cases in \figurename~\ref{mcneil2} and the penalty parameter is not too large, no constraint will be activated and the resulting model will be a GH. 
Suppose now the MLE is close enough to the case $\bgamma=\bzero$, with sufficiently small $\lambda$, large $\chi$, and $\psi$ close to zero. The low $||\bgamma||_{L_2}$ will make it advantageous to activate the constraint leading to symmetric models. The negative $\lambda$ will remove the third addend of the penalty, which is multiplied by $I(\lambda>0)$. For the second addend, the minimum among the three elements listed will be the third, as $\lambda$ at the MLE will definitely be much smaller than .5. Hence the penalty will essentially reduce to
$$
\frac{h}{4} \max\left( \frac{\left\|\bgamma\right\|_{L_2}}{\sqrt{d}},\left|\frac{1}{\lambda}\right|,|\psi|,\left|\frac{1}{\chi}\right|\right),
$$
and the $\max$ operator will lead all the constraints to activate ($\lambda\to -\infty$, $\psi\to 0$, $\chi\to\infty$, $||\bgamma||_{L_2}\to \bzero$), leading to the Gaussian model.

\section{Penalised maximum likelihood estimation}
\label{sec:Penalised maximum likelihood estimation}

%Let $\bPsi=\left\{\bmu,\bSigma,\bgamma,\lambda,\chi,\psi\right\}$ be the set of all the parameters of the GH distribution given in \eqref{eq:gh}.
We consider a penalised maximum likelihood (ML) approach, with the penalty term given in \eqref{60modelli} or \eqref{16modelli}, to estimate $\btheta$ in model~\eqref{eq:gh}.
%To find penalised maximum likelihood estimates of the parameters of the mixture model ... we need to maximize the penalised (observed data) 
Given both the random sample $S_n$ and a value for $h$, 
%Given a random sample $S_n=\left\{\bx_i; i=1,\ldots,n\right\}$ (observed data) from $\bX \sim \mathcal{GH}_d\left(\bmu,\bSigma,\bgamma,\lambda,\chi,\psi\right)$, 
the penalised ML estimation method is based on the maximization of the penalised (observed-data) log-likelihood function 
\begin{equation}
\ell_{\text{pen}}\left(\btheta|h\right)
= 
\sum_{i=1}^n \ln f\left(\bx_i;\btheta\right) - P_h\left(\bgamma,\lambda,\chi, \psi\right). 
\label{eq:GH penalised observed-data log-likelihood}
\end{equation}

However, the problem of directly maximising $\ell_{\text{pen}}\left(\btheta|h\right)$ over $\btheta$ is not particularly easy.
The penalised ML fitting is simplified considerably by the application of algorithms based on the expectation-maximization (EM) principle \citep{Demp:Lair:Rubi:Maxi:1977}.
These algorithms are the classical way to compute ML estimates for parameters of distributions which are defined as a mixture.

Regardless of the particular variant of the EM algorithm used, it is convenient to view the observed data as incomplete.
%Let $\Sa=\left\{\bX_i, i=1,\ldots,n\right\}$ be a sample of $n$ independent observations from the mixture model in ....
%In the context of the EM algorithm, $\Sa$ is considered incomplete.
The complete-data are $\left\{\left(\bx_i,w_i\right); i=1,\ldots,n \right\}$, where the missing variables $w_1,\ldots,w_n$ are defined -- based on the hierarchical representation given in \eqref{eq:hierarchical GH representation} -- so that
%$\bz_i=(z_{i1},\dots,z_{ik})'$ so that $z_{ij}=1$ if observation $i$ belongs to group $j$ and $z_{ij}=0$ otherwise.
%By saying that $\bZ_i$, random counterpart of $\vecz_i$, is distributed according to a multinomial distribution consisting of one draw from $k$ categories with probabilities $\bpi$, say $\bZ_i\sim\mathcal{M}_k\left(\bpi\right)$, the hierarchical representation for the mixture modeling framework is
$$
\bX_{i}|W_i=w_i \sim \mathcal{N}_d\left(\bmu + w_i \bgamma, w_i\bSigma\right),
$$
independently for $i\in \left\{1,\ldots,n\right\}$, and
$$
W_i  \sim     \mathcal{GIG}\left(\lambda,\chi,\psi\right). 
$$
%\begin{align}
%W_i  & \sim     \mathcal{GIG}\left(\lambda,\chi,\psi\right)  	\nonumber \\
%\bX_{i}|W_i=w_i & \sim \mathcal{N}_d\left(\bmu + w_i \bgamma , w_i\bSigma\right),
%\label{eq:mixture hierarchy}
%\end{align}
%where $W_{ij} \coloneqq W_{i}|z_{ij}=1$.
Because of this conditional structure, the penalised complete-data log-likelihood function can be written as
\begin{equation}
\ell_{\text{pen},c}\left(\btheta|h\right) = \ell_{1c}\left(\bmu,\bSigma,\bgamma\right)+\ell_{2c}\left(\lambda,\chi, \psi\right) - P_h\left(\bgamma,\lambda,\chi, \psi\right),
\label{eq:em2}
\end{equation}
%with $\bPsi=\left\{\left(\pi_j,\bmu_j,\bSigma_j,\bgamma_j,\theta_{j}\right); j=1,\ldots,K\right\}$, 
where 
\begin{align}
\ell_{1c}\left(\bmu,\bSigma,\bgamma\right) = & \sum_{i=1}^{n}\biggl[ -\frac{d}{2}\ln\left(2\pi\right)-\frac{d}{2}\ln\left(w_i\right)-\frac{1}{2}\ln|\bSigma|
-\frac{\delta\left(\bx_i;\bmu,\bSigma\right)   
}{2w_i}+ \nonumber \\
& + \left(\bx_i-\bmu\right)'\bSigma^{-1}\bgamma -
\frac{w_i}{2} \bgamma'\bSigma^{-1}\bgamma
\biggr],
\label{eq:em4}
\end{align}
and
\begin{equation}
\ell_{2c}\left(\lambda, \chi, \psi\right) = \sum_{i=1}^{n}\left\{
\left(\lambda-1\right) \ln \left(w_i\right)
- \frac{1}{2}  \frac{\chi}{w_i}
- \frac{1}{2} \psi w_i
- \frac{1}{2} \lambda \ln\left(\chi\right)
+ \frac{1}{2} \lambda  \ln\left(\psi\right)-\ln\left[2K_{\lambda}\left(\sqrt{\chi \psi}\right)\right]
\right\} .
\label{eq:ecme1}
\end{equation}
%\begin{align}
%\ell_{2c}\left(\lambda, \chi, \psi\right) = &\sum_{i=1}^{n}\left\{
%\left(\lambda-1\right) \ln \left(w_i\right)
%- \frac{1}{2}  \frac{\chi}{w_i}
%- \frac{1}{2} \psi w_i
%- \frac{1}{2} \lambda \ln\left(\chi\right)
%+ \frac{1}{2} \lambda  \ln\left(\psi\right)\right. \nonumber \\
%&\left.
%-\ln\left[2K_{\lambda}\left(\sqrt{\chi \psi}\right)\right]
%\right\} .
%\label{eq:ecme1}
%\end{align}
Working on $\ell_{\text{pen},c}\left(\btheta|h\right)$, we adopt the expectation-conditional maximization either (ECME) algorithm \citep{Liu:Rubi:TheE:1994}. 
%Working on $\ell_{\text{pen},c}\left(\btheta|h\right)$, the EM algorithm would iterate between two steps, one E-step and one M-step, until convergence.
%When using a penalization term we replace the M-step with the expectation-conditional maximization either (ECME) algorithm \citep{Liu:Rubi:TheE:1994}.
%we use the expectation-conditional maximization either (ECME) algorithm \cite{Liu:Rubi:TheE:1994}.     
The ECME algorithm is an extension of the expectation-conditional maximum (ECM) algorithm which, in turn, is an extension of the EM algorithm \citep{McLa:Kris:TheE:2007}. 
The ECM algorithm replaces the M-step of the EM algorithm by a number of computationally simpler conditional maximization (CM) steps.
The ECME algorithm generalizes the ECM algorithm by conditionally maximising on some or all of the CM-steps the incomplete-data (penalised) log-likelihood.
%As for the EM and ECM algorithms, the ECME algorithm monotonically increases the likelihood and reliably converges to a stationary point of the likelihood function \citep{McLa:Kris:TheE:2007}.
%Moreover, \citet{Liu:Rubi:TheE:1994} found the ECME algorithm to be nearly always faster than both the EM and ECM algorithms in terms of number of iterations, and that it can be faster in total computer time by orders of magnitude.
In our case, the ECME algorithm iterates between three steps, one E-step and two CM-steps, until convergence.
The two CM-steps arise from the partition of $\btheta$ as $\left\{\btheta_1,\btheta_2\right\}$, where $\btheta_1=\left\{\bmu,\bSigma\right\}$ and $\btheta_2=\left\{\bgamma,\lambda,\chi,\psi\right\}$.
The partition is chosen in such a way that all the parameters in the penalization function $P_h\left(\cdot\right)$ belongs to $\btheta_2$.
%\textcolor{blue}{The partition is chosen in such a way to have all the parameters impacted by the penalization in $\btheta_2$.}
%, with $\bmu=\left\{\bmu_j; j=1,\ldots,k\right\}$, $\bSigma=\left\{\bSigma_j; j=1,\ldots,k\right\}$, $\bgamma=\left\{\bgamma_j; j=1,\ldots,k\right\}$, $\blambda=\left\{\lambda_j; j=1,\ldots,k\right\}$, $\bchi=\left\{\chi_j; j=1,\ldots,k\right\}$, and $\bpsi=\left\{\psi_j; j=1,\ldots,k\right\}$. 

Below, we outline the generic iteration of the ECME algorithm. 
As in \cite{melnykov2018model,melnykov2019studying}, quantities/parameters marked with one dot will correspond to the previous iteration and those marked with two dots will represent the estimates at the current iteration. 

\subsection{E-Step}
\label{subsec:E-step}  

The E-step is only needed for the first CM-step of the algorithm -- where we update $\btheta_1$ -- and requires the calculation of 
\begin{equation}
Q\left(\btheta_1,\dot{\btheta}_2|\dot{\btheta}\right) = Q_1\left(\bmu,\bSigma,\dot{\bgamma}|\dot{\btheta}\right) + C, 
% + Q_3\left(\veclambda,\vecchi, \vecpsi|\dot{\bPsi}\right)
%-P_h\left(\bgamma,\veclambda,\vecchi, \vecpsi\right),
\label{eq:Q}
\end{equation}
the conditional expectation of $\ell_{\text{pen},c}\left(\btheta\left|h\right.\right)$ given the observed data, using the current fit $\dot{\btheta}$ for $\btheta$, with $\btheta_2$ fixed at $\dot{\btheta}_2$ and where $C$ is a constant not involving parameters inside $\btheta_1$.
In \eqref{eq:Q}, $Q_1\left(\bmu,\bSigma,\dot{\bgamma}|\dot{\btheta}\right)$ is the conditional expectation of $\ell_{1c}\left(\bmu,\bSigma,\bgamma\right)$ in \eqref{eq:em2}.

To compute $Q\left(\btheta_1,\dot{\btheta}_2|\dot{\btheta}\right)$ we need to replace any function $m\left(W_i\right)$ of the latent variable $W_i$ which appears in \eqref{eq:em4}, provided that it is related with either $\bmu$ or $\bSigma$, by $E_{\dot{\btheta}}\left[m\left(W_i\right)|\bX_i=\bx_i\right]$, where the expectation (as it can be noted by the subscript) is taken using the current fit $\dot{\btheta}$ for $\btheta$, $i=1,\ldots,n$.
In particular, 
%As concerns $W_{ij}$, it is only included in the Equations \eqref{eq:em4} and \eqref{eq:ecme1} and 
the functions satisfying these requirements, involved in \eqref{eq:em4}, are $m_1(w)=w$ and $m_2(w)=1/w$.
%, and $m_3(w)=\log(w)$.
To calculate the expectations of $m_1$ and $m_2$ we first note that 
$$
W_i|\bX_i=\bx_i \sim \mathcal{GIG}\left(\lambda -\frac{d}{2},\delta\left(\bx_i;\bmu,\bSigma\right)+ \chi,\bgamma'\bSigma^{-1}\bgamma +\psi\right).
$$
Therefore, according to \eqref{eq:ai} and \eqref{eq:bi}, respectively, we need to compute the following quantities 
\begin{align}
\dot v_i & \coloneqq \text{E}_{ \dot{\btheta}}\left(W_i|\bX_i=\bx_i\right) \nonumber \\
    & = \sqrt{\frac{\delta\left(\bx_i;\dot\bmu,\dot\bSigma\right)+ \dot\chi}{\dot\psi}}
		\frac{
		K_{\dot\lambda - \frac{d}{2}+1}\left\{\sqrt{\dot\psi \left[\delta\left(\bx_i;\dot\bmu,\dot\bSigma\right)+ \dot\chi\right]}\right\}
		}{
		K_{\dot\lambda -\frac{d}{2}} \left\{\sqrt{\dot\psi \left[\delta\left(\bx_i;\dot\bmu,\dot\bSigma\right)+ \dot\chi\right]}\right\}
		} \\
\dot u_i
& \coloneqq  \text{E}_{ \dot{\btheta}}\left(W_i^{-1}|\bX_i=\bx_i\right) \nonumber \\
& = \sqrt{\frac{\dot\psi}{\delta\left(\bx_i;\dot\bmu,\dot\bSigma\right) + \dot\chi}}
\frac{
K_{\dot\lambda - \frac{d}{2}+1}\left\{\sqrt{\dot\psi \left[\delta\left(\bx_i;\dot\bmu,\dot\bSigma\right) + \dot\chi\right]}\right\}
}{
K_{\dot\lambda - \frac{d}{2}} \left\{\sqrt{\dot\psi \left[\delta\left(\bx_i;\dot\bmu,\dot\bSigma\right) + \dot\chi\right]}\right\}
} -\frac{2\left(\dot\lambda -\frac{d}{2}\right)}{\delta\left(\bx_i;\dot\bmu,\dot\bSigma\right) + \dot\chi}.
%\nonumber \\ 
%l_i& \coloneqq  \text{E}_{ \dot{\btheta}}\left[\ln\left( W_i\right) |\bX_i=\bx_i\right]. 
\label{eq:Estep}
\end{align}
Then, by substituting $w_i$ with $\dot v_i$ and $1/w_i$ with $\dot u_i$ in $\ell_{1c}\left(\bmu,\bSigma,\bgamma\right)$, we obtain
\begin{equation}
Q_1\left(\bmu,\bSigma,\dot{\bgamma}|\dot{\btheta}\right) 
= 
\sum_{i=1}^{n}\left[
-\frac{1}{2}\ln|\bSigma|
-\frac{\dot u_i}{2}\delta\left(\bx_i;\bmu,\bSigma\right)   
+ \left(\bx_i-\bmu\right)'\bSigma^{-1}\dot\bgamma -
\frac{\dot v_i}{2} \dot\bgamma'\bSigma^{-1}\dot\bgamma
\right],
\label{eq:Q1}
\end{equation}
where we dropped the terms which are constant with respect to $\bmu$ and $\bSigma$.

\subsection{CM-step 1}
\label{subsec:CM-step 1}

The first CM-step requires the calculation of $\ddot\btheta_1$ as the value of $\btheta_1$ that maximizes $Q_1\left(\bmu,\bSigma,\dot{\bgamma}|\dot{\btheta}\right)$ in \eqref{eq:Q1}, with $\btheta_2$ fixed at $\dot\btheta_2$.
After simple algebra, we obtain the following updates 
\begin{align}
%\ddot{\pi}_j = &\frac{1}{n}\sum_{i=1}^n \hat{z}_{ij},\nonumber \\
%\ddot{\bgamma}_j =&\frac{\displaystyle \frac{1}{n_j}\sum_{i=1}^n \hat{z}_{ij} u_{ij} \left(\bar{\bx}_j - \bx_i \right)}{\bar{u}_{j}\bar{v}_{j}-1}, \nonumber \\
\ddot{\bmu} = & 
\displaystyle \frac{1}{n\dot{\bar{u}}}\left(\sum_{i=1}^n \dot u_i \bx_i - \dot{\bgamma}\right)
, \nonumber\\
\ddot{\bSigma} = & \left|\ddot{\bSigma}^* \right|^{-\frac{1}{d}}\ddot{\bSigma}^*
%\ddot{\theta}_j  = &\frac{\sum\limits_{i=1}^n \hat{z}_{ij}}{\sum\limits_{i=1}^n \hat{z}_{ij}\left(u_{ij}-1\right)},
\label{eq:m-step}
%&- \ddot{\bgamma} \left(\bar{\bx}_j-\ddot{\bmu}_j\right)'  + \bar{v}_j \ddot{\bgamma} \ddot{\bgamma}' . 
\end{align}
where 
\begin{equation}
\ddot{\bSigma}^* = \frac{1}{n}\sum_{i=1}^n  \dot u_i \left(\bx_i-\ddot{\bmu}\right) \left(\bx_i-\ddot{\bmu}\right)' - \left(\bar{\bx}-\ddot{\bmu}\right) \dot{\bgamma}' - \dot{\bgamma} \left(\bar{\bx}-\ddot{\bmu}\right)'  + \dot{\bar{v}} \dot{\bgamma} \dot{\bgamma}' ,
\label{eq:Sigma unconstrained update}
\end{equation}
$\dot{\bar{u}} = \sum_{i=1}^n \dot u_i/n$, $\dot{\bar{v}} = \sum_{i=1}^n \dot v_i/n$, and $\bar{\bx}= \sum_{i=1}^n \bx_i/n$.
In \eqref{eq:m-step}, the scalar $\left|\ddot{\bSigma}^* \right|^{-\frac{1}{d}}$ is needed to ensure the identifiability constraint $\left|\ddot{\bSigma} \right|=1$.

\subsection{CM-step 2}
\label{subsec:CM-step 2}

In the second CM-step, given $h$, we choose the value of $\btheta_2$ that maximizes $\ell_{\text{pen}}\left(\btheta|h\right)$ in 
\eqref{eq:GH penalised observed-data log-likelihood}, with $\btheta_1$ fixed at $\ddot\btheta_1$.
As a closed-form solution for $\ddot\btheta_2$ is not analytically available, numerical optimization is needed, and any general-purpose optimizer can be used with this aim.
Operationally, we perform an unconstrained maximization on $\real^{d+3}$, based on a ($\log$/$\exp$) transformation/back-transformation approach for $\chi$ and $\phi$, via the general-purpose optimizer \texttt{optim()} for \textsf{R}, included in the \textbf{stats} package.
In analogy with \citet{bagnato2021unconstrained}, we try two different commonly used algorithms for maximization: Nelder-Mead, which is derivatives-free, and BFGS which uses (numerical) second-order derivatives.
They can be passed to \texttt{optim()} via the argument \texttt{method}.
Once the two algorithms are run, we take the best solution in terms of $\ell_{\text{pen}}\left(\btheta|h\right)$; see, e.g., \citet{punzo2021multivariate} for a comparison of the two algorithms, in terms of parameter recovery and computational time, for ML estimation.
The choice to run both the algorithms is motivated by two facts: 1) sometimes the algorithms do not provide the same solution, and 2) it can happen that an algorithm does not reach convergence.
  
%\begin{equation}
%f\left(w_{ij}|\bx_i;\bmu_j,\bSigma_j,\bgamma_j,\lambda_j,\chi_j,\psi_j\right) = f_{\text{\tiny{GIG}}}\left(w_{ij};\lambda_j -\frac{d}{2},
%\delta\left(\bx_i;\bmu_j,\bSigma_j\right)+ \chi_j, 
 %\bgamma_j'\bSigma_j^{-1}\bgamma_j +\psi_j \right).
%%,\bgamma_j'\bSigma_j^{-1}\bgamma_j, \delta\left(\bx_i;\bmu_j,\bSigma_j\right)+2\theta_j ,1  \right),
%\label{eq:posterior}
%\end{equation}  

\subsection{Selecting the penalty parameter}
\label{sec:Selecting the penalty parameter}

The choice of the penalty parameter $h$ has got direct consequences on the estimation of $\btheta$ and, as a sub-product, on the selection of the best model in \figurename~\ref{mcneil2}.
As a data-driven method to select $h$,
%Among the data-driven methods for selecting the penalty parameter, 
we consider a simple grid-search partial leave-one-out likelihood cross-validation (LCV) strategy \citep{Ston:JRSSB:1974}; where the term ``grid-search'' refers to the fact that the LCV statistic is only evaluated on a convenient grid of values, while the term ``partial'' refers to the fact that we only allow to
a proportion $p$ of the sample to be left out one unit at a time.
These choices are motivated by the need to speed-up the computation that, otherwise, would be too much computationally cumbersome. 

In detail, we consider the LCV statistic
\begin{equation}
\text{LCV}_p\left(h\right) = \frac{1}{\left\lfloor pn\right\rfloor}\sum_{\bx_i \in S_{\left\lfloor pn\right\rfloor}} \ln\left[f\left(\bx_i;\widehat{\btheta}_{h,S_n\backslash \left\{\bx_i\right\}}\right)\right],
\label{eq:LCV statistic}
\end{equation}   
where 
%$S_n=\left\{\bx_i; i=1,\ldots,n\right\}$ is the whole sample of size $n$, 
$S_{\left\lfloor pn \right\rfloor}\subseteq S_n$ is the sub-sample, of size $\left\lfloor pn \right\rfloor$, which is allowed to be left out, and $\widehat{\btheta}_{ h, S_n \backslash \left\{\bx_i\right\} }$ is the penalised ML estimate of $\btheta$, with penalty parameter $h$, obtained on $S_n\backslash \left\{\bx_i\right\}$ (refer to Section~\ref{sec:Penalised maximum likelihood estimation}).
For each value of $h$ in a pre-specified grid $G$, we first compute $\text{LCV}_p\left(h\right)$; then, we select the value of $h$ in correspondence to the maximum value of this statistic.

\section{Simulation study}
\label{sec:Simulation study}

In this section, we describe the results of a simulation study conducted with the aim of investigating the ability of our multiple-choice LASSO procedure in discovering the true data generating model (DGM) among those in \figurename~\ref{mcneil2}.

For each of the following DGMs we consider 50 randomly generated datasets, of size $n=1000$, with $d=2$ dimensions.
The DGMs considered are: normal (N), $t$, Cauchy (C), Laplace (L), symmetric generalised hyperbolic (SGH), skew-$t$ (S$t$), variance gamma (VG), and asymmetric Laplace (AL).  
The DGMs share the same location parameter $\bmu=\bzero$ and scale matrix $\bSigma=\bI$, with $\bI$ denoting the identity matrix.
We fix $\bgamma=\left(-0.5,0.8\right)'$ for the skewed DGMs (S$t$, VG, and AL).
Parameters $\lambda$, $\chi$, and $\psi$ vary according to the considered DGM; \tablename~\ref{tab:true parameters} provides the precise values of these parameters for each.

\begin{table}[!ht]	
\centering																				
% \resizebox{\textwidth}{!}{
\begin{tabular}{lcd{3.3}d{3.3}d{3.3}d{3.3}d{3.3}d{3.3}}
 %\begin{tabular}{lr*{6}{>{\centering\arraybackslash}p{.075\linewidth}}}
%{|*{9}{C|}}
%{lc rrrrrrrrr} 
%{lc@{\hskip 0.3in}r@{\hskip 0.3in}rr@{\hskip 0.3in}rrc}	
%{lr*{9}{p{1cm}}}%
\toprule													
	&		&		\multicolumn{6}{c}{DGM} \\	
\cline{3-8}																			
Parameter	& &		\multicolumn{1}{c}{\text{N}}	&		\multicolumn{1}{c}{\text{$t$, S$t$}}	&		\multicolumn{1}{c}{\text{C}} 	&		\multicolumn{1}{c}{\text{L, AL}} 	&		\multicolumn{1}{c}{\text{SGH}}	& \multicolumn{1}{c}{\text{VG}}	  \\	
\midrule																																		
$\lambda$	&	& -20	&		-1	    &		-0.5  &		1  	&		-1	 	&		1.5	   	\\	
$\chi$	  &	&	100	  &		2	      &		   2	&		0.001	&		2	 	&		0.001  	 	\\
$\psi$	  &	&	0.001	&		0.001		&	 0.001	&		0.5	    &		3	 	&		0.5	  	 	\\
\bottomrule
\end{tabular}			
%}																		
\caption{Parameters $\lambda$, $\chi$ and $\psi$ of the DGMs used in the simulation study.}		
\label{tab:true parameters}																			
\end{table}	

We use our penalised ML procedure on each generated dataset. 
We select the penalty parameter $h$ with the LCV strategy described in Section~\ref{sec:Selecting the penalty parameter},
using the grid 
\linebreak 
$G=\left\{0, 5, 10, 15, 20, 25, 30, 35, 40, 45, 50, 60, 70, 80, 100\right\}$ and a proportion $p=0.1$ of observations which are allowed to be left out one at a time.
  
\tablename~\ref{tab:Simulation Results} shows the number of times our multiple-choice LASSO method selects each model in our family of models.
Here, there are some models that are fitted to the data but they are not used as DGMs; these models are the normal-inverse
Gaussian (NIG), hyperbolic (H), hyperbolic univariate marginals (HUM), symmetric normal-inverse
Gaussian (SNIG), symmetric variance gamma (SVG), symmetric hyperbolic (SH), skew-Cauchy (SC), and generalized
hyperbolic (GH).
Results are organised as a contingency table where the true DGM is given by column and the models in the GH-family by row.
The shadowed cells report the true positive count (TPC), measuring the number of times over the replicates that the multiple-choice LASSO approach discovers the true DGM.
We can note how, regardless of the DGM, our approach is able enough to recognize the true underlying DGM, being the counts mainly concentrated on the shadowed cells.
The best results are obtained for the $t$-DGM, where the TCP is the maximum possible (50).
On the opposite side, the worst results are obtained for the N-DGM, where $\text{TCP}=42$; in the remaining 8 cases, the more general skew-$t$ distribution is selected.
%; however, for some DGMs, such as the N, their considerable similarity makes discovering the true model more difficult.
%%assessment of the true model challenging.
%This is probably due to the considered parameterizations under the MSVG and MSH DGPs.
%%being the modal choice always toward the true model.
%%The last column of these tables helps us to view the overall results by showing
%The last column of each table shows the average number of times a model different from the underlying DGP is selected; this can be meant as a sort of false positive percentage (FPP). 
%The ideal situation for AIC and BIC should have zero off-diagonal values.
\begin{table}[!ht]	
\centering																				
% \resizebox{\textwidth}{!}{
 \begin{tabular}{lc*{8}{>{\centering\arraybackslash}p{.065\linewidth}}}
%{|*{9}{C|}}
%{lc rrrrrrrrr} 
%{lc@{\hskip 0.3in}r@{\hskip 0.3in}rr@{\hskip 0.3in}rrc}	
%{lr*{9}{p{1cm}}}%
\toprule													
	&		&		\multicolumn{8}{c}{DGM}																																		\\	
\cline{3-10}																			
Fitted	&		&		N	&		$t$	&		C 	&		L 	&		SGH	&		S$t$ &		AL	&		VG	 	 \\	
\midrule																																		
N	  &	&\cellcolor{lightgray}	42	&		0	&		0	&		0	&		0	&		0	&		0	&		0  	\\	
$t$	&	&	0	&		\cellcolor{lightgray} 50	&		1	&		0	&		4	&		0	&		0	&		0 	 	\\
C	&	&	0	&		 0	&		\cellcolor{lightgray} 49	&		0	&		0	&		0	&		0	&		0 	 	\\
L	&	&	0	&		 0	&		 0	&		\cellcolor{lightgray} 46	&		0	&		0	&		0	&		0 	 	\\
SGH	&	&	0	&		 0	&		 0	&	0	&			\cellcolor{lightgray}  45	&		0	&		0	&		0 		\\
S$t$	&	&	8	&		 0	&		 0	&	0	&			  0	&	\cellcolor{lightgray}	49 &		0	&		0	 	 	\\
AL	&	&	0	&		 0	&		 0	&	0	&			 0	&		0 &	\cellcolor{lightgray} 	44 &		0	 	 	\\
VG	&	&	0	&		 0	&		 0	&	0	&			  0	&		0 &		3	&	\cellcolor{lightgray}	48	 	\\
NIG	&	&	0	&		 0	&		 0	&	0	&			  0	&		0 &		0	&		0	  	\\
H	  &	&	0	&		 0	&		 0	&	0	&			  0	&		0 &		0	&		1	 	 	\\
HUM &	&	0	&		 0	&		 0	&	1	&			  0	&	  0 &		0	&		0	  \\
SNIG&	&	0	&		 0	&		 0	&	0	&			  0	&		0	&		0	&		  0  	\\
SVG	&	&	0	&		 0	&		 0	&	3	&			  1	&		0	&		0	&		  0 	\\
SH	&	&	0	&		 0	&		 0	&	0	&			  0	&		0	&		0	&		  0  \\
SC	&	&	0	&		 0	&		 0	&	0	&			  0	&		  1 &		0	&		0	  \\
GH	&	&	0	&		 0	&		 0	&	0	&			  0	&		  0 &		3	&		1	  	\\
\bottomrule
\end{tabular}			
%}
\caption{Number of times the multiple-choice LASSO approach selects each model.
The true DGM is shown by column, while the models in the GH-family are given by row.}		
\label{tab:Simulation Results}																			
\end{table}	

\section{Concluding remarks}
\label{sec:conclusions}

In this work we have put forward a taxonomy of the GH family, and showed how one can perform simultaneous estimation and selection of nested models within the family.
We argue that the GH family is flexible enough to fit well a wide range of distributions in real applications, and that the model selection procedure is effective in providing a simple and interpretable model class without sacrificing goodness of fit.
We also have introduced the multiple-choice LASSO. We believe
adaptive choice of the shape parameters within the GH family is only one of the possible applications of the multiple-choice LASSO,
and that its theoretical properties deserve further investigation. Additionally, there are other flexible and general parametric families of
distributions that might benefit from an approach similar to the one proposed in this work (e.g., \citealp{gera:farc:20}).

\appendix

\section*{Appendix}

\section{Generalised inverse Gaussian distribution}
\label{app:Generalised inverse Gaussian distribution}

The random variable $W$ has a generalised inverse Gaussian (GIG) distribution if its pdf is
\begin{equation}
f_{\text{\tiny{GIG}}}\left(w;\lambda,\chi,\psi\right)=
\left(\frac{\psi}{\chi}\right)^{\frac{\lambda}{2}}\frac{w^{\lambda-1}}{2 K_\lambda\left(\sqrt{\psi \chi}\right)} \exp\left[-\frac{1}{2}\left(\psi w+\frac{\chi}{w}\right)\right], \qquad w>0,
\label{eq:GIG pdf}
\end{equation}
where the parameters satisfy the conditions: $\chi>0$ and $\psi \geq 0$, if $\lambda<0$; $\chi>0$ and $\psi>0$, if $\lambda=0$; $\chi\geq 0$ and $\psi>0$, if $\lambda>0$.
If $W$ has the pdf in \eqref{eq:GIG pdf}, then we simply write $W \sim \mathcal{GIG}\left(\lambda,\chi,\psi\right)$. 
%The GIG distribution was introduced by \cite{good1953population} and has several special cases as, for example, the gamma ($\chi=0$, $\lambda>0$) and the inverse Gaussian ($\lambda=-1/2$) distributions; for further details and properties see \cite{barndorff1977}, \cite{blaesild1978shape}, and \cite{halgreen1979self}. 
The expectations of $W$ and $1/W$, used in Section~\ref{subsec:E-step}, are
\begin{equation}
\text{E}\left(W\right)=\sqrt{\frac{\chi}{\psi }}\frac{K_{\lambda+1}(\sqrt{\psi \chi})}{K_{\lambda}(\sqrt{\psi \chi})}
\label{eq:ai}
\end{equation}
and
\begin{equation}
\text{E}\left(\frac{1}{W}\right)=\sqrt{\frac{\psi }{\chi}}\frac{K_{\lambda+1}(\sqrt{\psi \chi})}{K_{\lambda}(\sqrt{\psi \chi})}-\frac{2\lambda}{\chi}.
\label{eq:bi}
\end{equation}
%and
%\begin{equation}
%\mathbb{E}\left(\log W\right)=\log\left(\sqrt{\frac{\chi}{\psi }}\right)+\frac{1}{K_{\lambda}(\sqrt{\psi \chi})}\frac{\partial}{\partial \lambda}K_{\lambda}(\sqrt{\psi \chi}).
%\label{eq:ci}
%\end{equation}

%\begin{align}
 %\text{E} \left(W^r\right) &= \left(\frac{\chi}{\psi}\right)^{\frac{r}{2}}
%\frac{
%K_{\lambda+r}\left(\sqrt{\chi\psi} \right)
%}{
%K_{\lambda}\left(\sqrt{\chi\psi} \right)
%}, \quad r \in  \mathbb{R} \label{eq:expectGIG} \\
%\text{E} \left[\ln\left(W\right)\right] & = \left.\frac{\partial  \text{E} \left(W^r\right)}{\partial r}\right|_{r=0}.
%\label{eq:expectGIGlog}
%\end{align}

\section{Special and limiting cases of the GH distribution}
\label{app:Special and limiting cases of the GH distribution}

\subsection{GH $\rightarrow$ Skew-$t$ $\rightarrow$ $t$ $\rightarrow$ Gaussian}
\label{app:skewt}

If $\lambda<0$ and $\psi \rightarrow 0$, then $W \sim \mathcal{GIG}\left(\lambda,\chi,\psi\right)$ tends to $W \sim \mathcal{IG}\left(-\lambda, \frac{\chi}{2}\right)$, where $\mathcal{IG}\left(\cdot\right)$ denotes the inverse gamma distribution.  
%the GIG distribution in \eqref{eq:GIG pdf} becomes the so-called inverse gamma distribution, in symbols $W \sim IG\left(-\lambda, \frac{\chi}{2}\right)$.
Therefore, the NMVM representation in \eqref{eq:vmm} becomes
\begin{equation*}
\bX 
%&=\bmu+W\bgamma+\sqrt{W}\vecU \\
%&
=\bmu-V\frac{\chi}{2\lambda} \bgamma+\sqrt{V}\bar{\vecU},
\end{equation*}
where $V=-\frac{2\lambda}{\chi}W \sim \mathcal{IG}\left(-\lambda, -\lambda\right)$ and $\bar{\vecU} \sim\mathcal{N}_d\left(\bzero,-\frac{\chi}{2\lambda}\bSigma\right)$, with $\left|\bSigma\right|=1$.
Note that, thanks to the multiplicative factor $-\chi/\left(2\lambda\right)$, $\left|\text{Cov}\left(\bar{\vecU}\right)\right|=\left[-\chi/\left(2\lambda\right)\right]^d\left|\bSigma\right|=\left[-\chi/\left(2\lambda\right)\right]^d$ can be any positive real number. 
Under this setting, $\bX\sim \mathcal{S}t_{d}\left(\bmu,-\frac{\chi}{2\lambda}\bSigma,-\frac{\chi}{2\lambda}\bgamma,-2\lambda\right)$, which represents a skew-$t$ distribution with location parameter $\bmu$, scale matrix $-\frac{\chi}{2\lambda}\bSigma$, skewness parameter $-\frac{\chi}{2\lambda}\bgamma$, and $\nu=-2\lambda$ degrees of freedom \citep{hu2005calibration,murray2014mixtures}.
Compared to the GH-parametrization adopted by \citet{mcnicholas16a}, in our case, because of the identifiability constraint $\left|\bSigma\right|=1$, there is no reason to force $\chi$ and $\lambda$ to be related as $\chi=\nu=-2\lambda$.
In other words, with our parametrization, $\chi$ is unconstrained.
%free to be any positive real number.  
%Note that, due to the constraint $\left|\bSigma\right|=1$, $\chi$ is free to vary on the positive real line.
Indeed, if we impose the constraint $\chi=\nu=-2\lambda$ with our parametrization, then we would get $\left|\text{Cov}\left(\bar{\vecU}\right)\right|=1$.
% the covariance matrix of $\bar{\vecU}$ will be restricted to have determinant equal to 1.
If, in addition, $\bgamma=\bzero$, then $\bX\sim t_{d}\left(\bmu,-\frac{\chi}{2\lambda}\bSigma,-2\lambda\right)$, which represents a $t$ distribution with location parameter $\bmu$, scale matrix $-\frac{\chi}{2\lambda}\bSigma$, and $\nu=-2\lambda$ degrees of freedom.
Finally, if we further consider $\lambda =-\chi/\left(2c\right)$, with $c>0$, and $\chi\rightarrow \infty$, then we obtain $\bX \sim\mathcal{N}_d\left(\bzero,c \bSigma\right)$ as a limiting case.  
%Moreover, $\bX \sim\mathcal{N}_d\left(\bzero,c \bSigma\right)$ when $\lambda =-\frac{\chi}{2c}$, with $c>0$, and $\chi\rightarrow \infty$. 
%Also here, it is easy to observe that $\chi$ should not be restricted to assume the value $-2\lambda$ since in this way only normal distributions with covariance matrices having determinant equal to 1 will be obtained.  

\subsection{GH $\rightarrow$ Variance Gamma $\rightarrow$ Asymmetric Laplace $\rightarrow$ Laplace}

If $\lambda>0$ and $\chi \rightarrow 0$, then $W \sim \mathcal{GIG}\left(\lambda,\chi,\psi\right)$ tends to $W \sim \mathcal{G}\left(\lambda,\frac{\psi}{2}\right)$, where $\mathcal{G}\left(\cdot\right)$ denotes the gamma distribution.
Then, the NMVM representation in \eqref{eq:vmm} becomes
\begin{equation*}
\bX
%&=\bmu+W\bgamma+\sqrt{W}\vecU \\
%&
= \bmu+V\frac{\psi}{2\lambda} \bgamma+\sqrt{V}\bar{\vecU},
\end{equation*}
where $V=\frac{2\lambda}{\psi}W \sim \mathcal{G}\left(\lambda, \lambda\right)$ and $\bar{\vecU} \sim\mathcal{N}_d\left(\bzero,\frac{\psi}{2\lambda}\bSigma\right)$, with $\left|\bSigma\right|=1$.
Note that, thanks to the multiplicative factor $\psi/\left(2\lambda\right)$, $\left|\text{Cov}\left(\bar{\vecU}\right)\right|=\left[\psi/\left(2\lambda\right)\right]^d\left|\bSigma\right|=\left[\psi/\left(2\lambda\right)\right]^d$ can be any positive real number. 
Under this setting, $\bX\sim \mathcal{VG}_{d}\left(\bmu,\frac{\psi}{2\lambda}\bSigma,\frac{\psi}{2\lambda}\bgamma,\lambda\right)$, which represents a variance gamma distribution with location parameter $\bmu$, scale matrix $\frac{\psi}{2\lambda}\bSigma$, skewness parameter $\frac{\psi}{2\lambda}\bgamma$, and shape parameter $\lambda$ \citep{nitithumbundit2020ecm}.
Compared to the VG-parametrization adopted by \citet{nitithumbundit2020ecm} and \citet{mcnicholas16a}, in our case, because of the identifiability constraint $\left|\bSigma\right|=1$, there is no reason to force $\psi$ and $\lambda$ to be related as $\psi=2\lambda$.
In other words, with our parametrization, $\psi$ is unconstrained.
Indeed, if we impose the constraint $\psi=2\lambda$ with our parametrization, then we would get $\left|\text{Cov}\left(\bar{\vecU}\right)\right|=1$.
If, in addition, $\lambda=1$, then $V \sim \mathcal{E}\left(1\right)$, which is a standard exponential distribution, and $\bX\sim \mathcal{AL}_{d}\left(\bmu,\frac{\psi}{2}\bSigma,\frac{\psi}{2}\bgamma\right)$, which represents an asymmetric Laplace distribution with location parameter $\bmu$, scale matrix $\frac{\psi}{2}\bSigma$, and skewness parameter $\frac{\psi}{2}\bgamma$; see \citet{kozubowski2000multivariate} and \citet{Morr:Punz:McNi:Brow:Asym:2018}. 
Finally, if we further consider $\bgamma=\bzero$, then $\bX\sim \mathcal{L}_{d}\left(\bmu,\frac{\psi}{2}\bSigma\right)$, which represents a Laplace distribution with location parameter $\bmu$ and scale matrix $\frac{\psi}{2}\bSigma$; see \citet{kozubowski2000multivariate}.  

\subsection{GH $\rightarrow$ Normal-Inverse Gaussian $\rightarrow$ Skew-Cauchy $\rightarrow$ Cauchy}

If $\lambda=-1/2$, then $\bX\sim \mathcal{NIG}_{d}\left(\bmu, \bSigma, \bgamma, \chi, \psi\right)$, which denotes the normal-inverse Gaussian distribution with location parameter $\bmu$, scale matrix $\bSigma$, skewness parameter $\bgamma$, and concentration parameters $\chi$ and $\psi$ \citep{o2016clustering}. 
If, in addition, $\psi \rightarrow 0$, then $\bX\sim \mathcal{SC}_{d}\left(\bmu, \chi \bSigma, \chi \bgamma\right)$, which represents the skew-Cauchy distribution with with location parameter $\bmu$, scale matrix $\chi \bSigma$, and skewness parameter $\chi \bgamma$ \citep{cabral2012multivariate}.
Note that, $\mathcal{SC}_{d}\left(\bmu, \chi \bSigma, \chi \bgamma\right)$ can be also obtained as a special case of $\mathcal{S}t_{d}\left(\bmu,-\frac{\chi}{2\lambda}\bSigma,-\frac{\chi}{2\lambda}\bgamma,-2\lambda\right)$ when $\lambda=-1/2$; refer to Section~\ref{app:skewt}. 
% from the skew-$t$ in Section \ref{app:skewt} when $\lambda=-1/2$ is selected.
Finally, if we further consider $\bgamma=\bzero$, then $\bX\sim \mathcal{C}_{d}\left(\bmu,\chi \bSigma\right)$, which represents a Cauchy distribution with location parameter $\bmu$ and scale matrix $\chi\bSigma$.
%; see \citet{kozubowski2000multivariate}.  

\bibliographystyle{Chicago}
\bibliography{biblio}

\end{document}